# Shared control schematic for brain controlled vehicle based on fuzzy logic

Na Dong   Wen-qi Zhang and Zhong-ke Gao*

School of Electrical and Information Engineering, Tianjin University, Tianjin, China

**ABSTRACT:** Brain controlled vehicle refers to the vehicle that obtains control commands by analyzing the driver's EEG through Brain-Computer Interface (BCI). The research of brain controlled vehicles can not only promote the integration of brain machines, but also expand the range of activities and living ability of the disabled or some people with limited physical activity, so the research of brain controlled vehicles is of great significance and has broad application prospects.

At present, BCI has some problems such as limited recognition accuracy, long recognition time and limited number of recognition commands in the process of analyzing EEG signals to obtain control commands. If only use the driver's EEG signals to control the vehicle, the control performance is not ideal. Based on the concept of Shared control, this paper uses the fuzzy control (FC) to design an auxiliary controller to realize the cooperative control of automatic control and brain control. Designing a Shared controller which evaluates the current vehicle status and decides the switching mechanism between automatic control and brain control to improve the system control performance. Finally, based on the joint simulation platform of Carsim and MATLAB, with the simulated brain control signals, the designed experiment verifies that the control performance of the brain control vehicle can be improved by adding the auxiliary controller.

## 1 Introduction

The brain is the control center of all movement and language in the human body, and sends instructions to the body through the external nerve as the medium[1]. However, some unexpected factors may cause such diseases as cerebral palsy and spinal cord injury, so the brain can't communicate with the outside world through the normal neuromuscular system. These disabled people have normal brains, can think normally but can't take care of themselves, have limited mobility, suffer physical pain, are under great psychological pressure, and are very dependent on others. As a special group in society, these disabled people often cause various social problems.

It is in this context that the brain-computer interface (BCI) emerged, which is an information communication channel directly established between the brain and the computer or peripheral devices without relying on the conventional output pathway of the brain[2]. After sampling and analyzing human brain electrical signals, the idea of human control can be obtained, and then the idea of control can be exported to computers or peripherals, and finally the control can be realized. The term BCI was first coined in the 1970s by Jacques Vidal. In 1999, in the United States, the first session of brain-computer interface symposium was held, which had a clear definition of brain-computer interface[3]. Currently, BCI has two methods to obtain brain electrical signals, invasive and non-invasive, while non-invasive BCI is more widely used because it does not require surgery to implant the acquisition device into the brain. The non-invasive brain-computer interface system generally consists of three parts: signal acquisition, signal decoding and external control equipment. The structural schematic diagram is shown in figure 1.

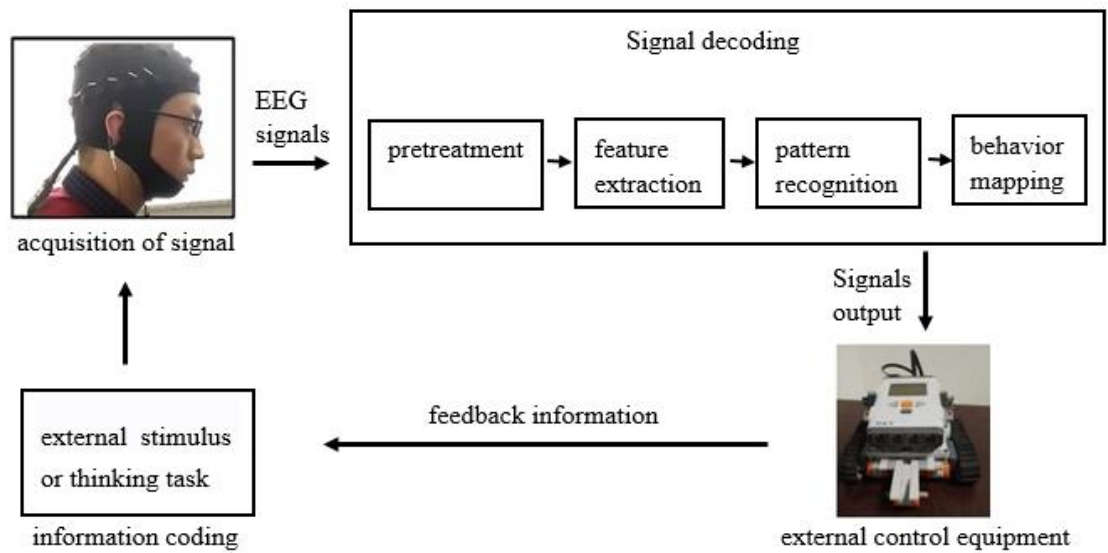

FIG. 1 schematic diagram of brain-computer interface (BCI) structure

Recently, many researchers have carried out researches on brain-computer interface technology, hoping to realize the control of external devices through BCI. Some achievements have been made, such as brain controlled wheelchair[4][5], brain controlled vehicle[6]-[8], and brain controlled robot[9]-[12]. In many applications of BCI, brain controlled vehicle refers to obtaining driving command through analyzing human brain electrical signals through BCI to control the vehicle, and the driver controls the movement of the vehicle through BCI without using the limbs [13].

Research on brain-controlled vehicles has provided a new way of driving for disabled people, greatly expanding their range of activities and improving their living standards and self-care ability. It also provides valuable experience for other brain controlled equipment. As a new way of driving, brain controlled driving can also promote the study of human-computer interaction and more intelligent equipment. So the research on brain-controlled vehicles is significant.

At present, the research on brain controlled vehicles at home and abroad is still in its infancy, and the main research contents can be divided into two parts. First part, direct control of vehicle movement by BCI. In other words, BCI is used to analyze the driver's EEG signal and output the obtained driving intention directly to the vehicle as the control signal to realize the control. However, such online control requires BCI to have ultra-short instruction time and high recognition accuracy. But the current performance of BCI is still not up to the requirements. In the second part, in the current situation of limited BCI performance, combined with advanced control methods, a Shared control method is adopted to combine brain-computer interface technology and intelligent control technology. The Shared control method is adopted to improve the control performance of brain controlled vehicles.

In the case of the limited function of BCI, the performance of the car which is controlled only by EEG signals is poor. Based on the concept of Shared control, this paper uses fuzzy control (FC) to design the auxiliary controller to realize the collaborative control between automatic control and brain control. Designing a Shared controller which evaluates the current vehicle status and decides the switching mechanism between automatic control and brain control to improve the system control performance. Finally, based on the joint simulation platform of Carsim and MATLAB, with the simulated brain control signals, the designed experiment verifies that the control performance of the brain control vehicle can be improved by adding the auxiliary controller.

The research content of this paper is mainly divided into the following aspects:

1) establish auxiliary controller based on fuzzy control theory

2) based on the concept of Shared control, design a Shared controller which evaluates the current vehicle status and determines the switching mechanism between automatic control and brain control, including the simple realized switching according to the error size and the optimized switching according to the cost function.

3) simulated the brain control signal in MATLAB, which is used to control the vehicle in Carsim

4) finally, based on the joint simulation platform of Carsim and MATLAB, compare the control performance of the vehicle controlled only by brain signals and the vehicle controlled by brain signals with an auxiliary controller based on Shared control.

## 2 The Shared control system of brain-controlled vehicles

The idea of Shared control is: both of people and intelligent controllers control the system together and complement each other's advantages. In other words, people's adaptive ability and intelligence are integrated with the precision of machines. The purpose of Shared control is to combine human intelligence with high efficiency of machine, so as to reduce human workload and ensure the reliability and stability of machine control. Research on Shared control has a history of recent years, but there is no unified definition at present, and there are many ways to propose, such as remote control, quasi-autonomous control, sharing and exchange control[14].

Now, the Shared control of brain controlled vehicles can be divided into two types: the Shared control at the decision-making level and the Shared control at the execution level[13]. In the first kind of brain controlled vehicle system, people, as a high-level decision maker, rely on BCI to output their intention to the vehicle, and then implement it automatically through the existing automatic control strategy. In the second kind of brain controlled vehicle system, an auxiliary controller is designed through some automatic control technologies. People control the vehicle together with this controller.

In the process of the entire brain controlled driving, driver relies on the vehicle state and environmental information to make decision (e.g., the steering wheel angle)and outputs control instruction through BCI. At the same time auxiliary controller based on fuzzy logic reasoning also outputs a control command. The shared controller constantly comprehensives the output of both and the status of vehicle to determine the switching mechanism between automatic control and brain control. Finally, realize the control of vehicles. The entire brain controlled driving process is a closed - loop system of people and vehicles. The block diagram of the system is as follows:

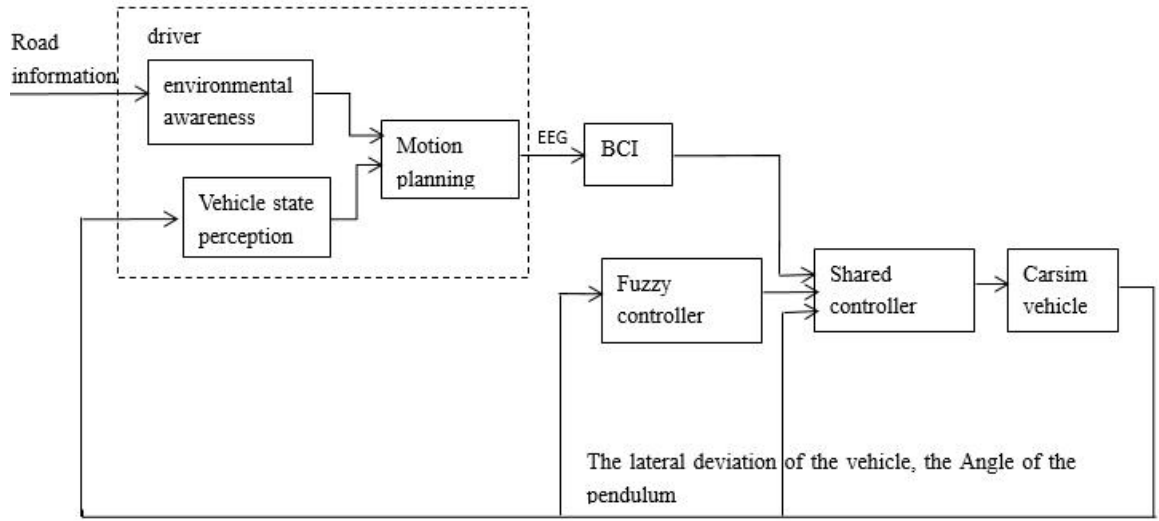

FIG. 2 overall system diagram of brain-controlled vehicles

## 3 Auxiliary fuzzy controller design

Fuzzy control is based on fuzzy set theory and fuzzy logic reasoning[15]. Fuzzy control treats the control object as a black box, and changes people's operational experience and relevant knowledge into some rules of language expression, and then makes the executing mechanism operate on the black box according to these rules, so it is not dependent on the system's transfer function. Because fuzzy rules are designed by people's experience and knowledge, just like offline human reasoning decisions, human intelligence becomes part of the control system. This is in line with the idea of brain control, so this paper chooses to design an auxiliary controller based on fuzzy control to help brain control complete the control of vehicles.

Because the structure of Mamdani fuzzy controller is simple and fast, this paper selects Mamdani fuzzy controller, whose structure is shown in figure 3.

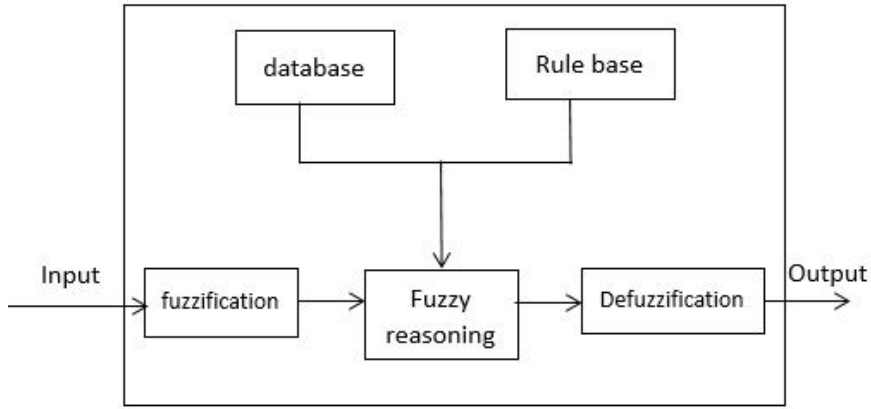

FIG. 3 Mamdani fuzzy controller structure

The process of establishing the fuzzy controller in this paper is as follows:

1) The determined input variables are the deviation e of the vehicle from the center line of the road, the change rate de of the deviation, and the output variables are the control variable u (steering wheel angle of the vehicle). The fuzzy sets of all three are {NB,NM,NS,ZO,PS,PM,PB}.

2) Determine the membership function of input and output

When choose the membership function of input variables e, de and output variable u, choose the function whose middle range is relatively narrow, both sides are wide. So that when error is small it can soon be stable. Besides, choose the membership function to cross and leave no space. As for shape, e and u choose relatively steep triangle shape with the combination of relatively smooth gaussian shape and de whole choose relatively smooth gaussian shape. Its graphics as shown in the figure 4

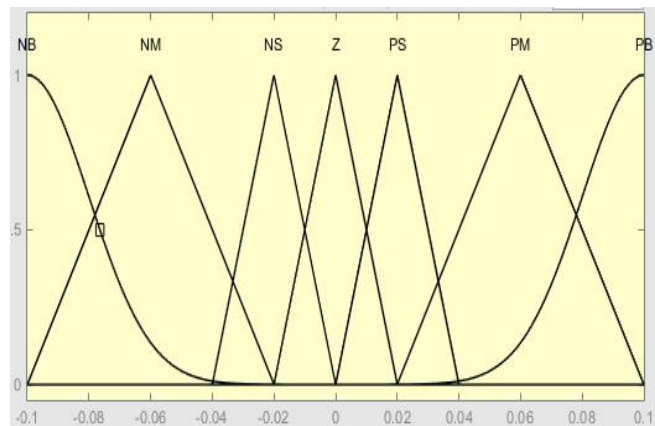

(a) membership function of input variable

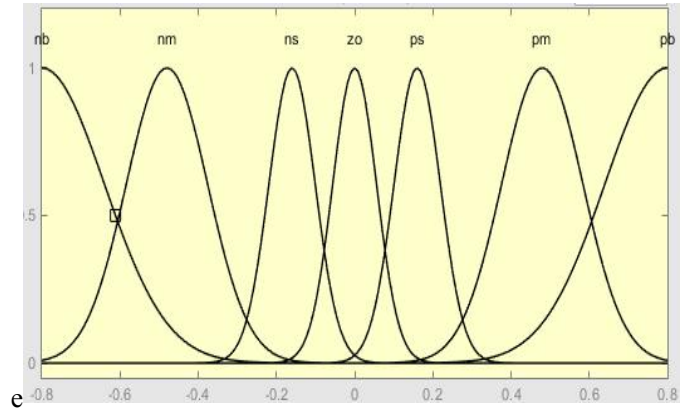
(b) membership function of input variable de

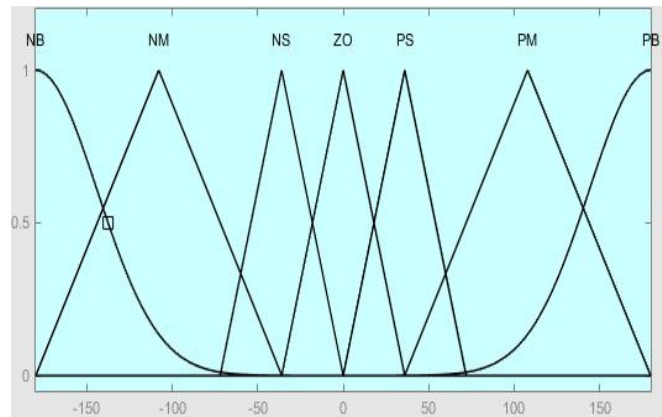
(c) membership function of output variable u

Figure 4. Membership function of input and output variables

3) Establish fuzzy control rules

Generally, fuzzy control rules can be obtained in two ways: by summarizing the experience and knowledge of experts, and by summarizing the practical operation process skillfully. In this paper, when selecting fuzzy inference rules, a set of 49 control rules are obtained by referring to the data. After that, some of them are modified according to the simulation experiment process. The control rules are as follows:

If e is NB and de is NB then u is NB

If e is NM and de is NB then u is NB

If e is NS and de is NB then u is NM

…

If e is PM and de is PB then u is PB

If e is PB and de is PB then u is PB

4) Establish the fuzzy control table

The fuzzy control table can be made from the fuzzy rules used in step 3 as follows

Table 1 fuzzy control table

| e \ de | NB | NM | NS | ZO | PS | PM | PB |
|---|---|---|---|---|---|---|---|

| NB | NB | NB | NM | NM | NS | NS | ZO |
|----|----|----|----|----|----|----|----|
| NM | NB | NM | NM | NS | NS | ZO | PS |
| NS | NM | NM | NS | NS | ZO | PS | PS |
| ZO | NM | NS | NS | ZO | PS | PS | PM |
| PS | NS | NS | ZO | PS | PS | PM | PM |
| PM | NS | ZO | PS | PS | PM | PM | PB |
| PB | ZO | PS | PS | PM | PM | PB | PB |

5) Fuzzy reasoning. By searching the fuzzy control table in step 4 online, the complexity of operation can be reduced when calculating the output value. As for the synthesis operation, the synthesis method of take large - take small is adopted.

6) Determine the method of anti-blurring: the center of gravity method is adopted as the method of anti-blurring.

## 4 Shared controller design

Shared control is a kind of control method integrating intelligent control method and people's own intention. A typical Shared controller model is a black box with two inputs and one output[14]. According to the environmental information and state of the vehicle, the driver makes decision. Then BCI outputs the decision command. At the same time, auxiliary fuzzy controller based on fuzzy logic reasoning also outputs a control command. The shared controller constantly comprehensives the output of both and the status of vehicle to determine the switching mechanism between automatic control and brain control. Finally, realize the control of vehicles. Therefore, the shared controller designed in this paper is a black box with three inputs and one output.

Since the research of the brain controlled vehicle in this paper mainly focuses on the lateral movement of the brain controlled vehicle, the three inputs of the designed Shared controller are respectively set as brain control signal, the output of the auxiliary fuzzy controller and the lateral deviation of the vehicle. The function of the Shared controller is to determine the switching mechanism between brain control and automatic control. For the setting of switching mechanism, two schemes are proposed in this paper.

The first:

It can be seen from the shared control scheme proposed in chapter 2 that the driver's driving intention is respected when the driving deviation is not large. When the deviation is large, the auxiliary controller intervenes to ensure the control precision of the vehicle and reduce the burden of the driver. Therefore, the mechanism that determines the switching in the Shared controller is the magnitude of error. By setting a threshold value, the shared controller switches to brain control when error is less than the threshold value, and switches to auxiliary fuzzy control when error is greater than the threshold value. By the principle of fuzzy control which is introduced before, the fuzzy control is adjusted according to the set of fuzzy rules based on the error size, no control action when no difference. If choose maximum value of deviation e which is the input variable of fuzzy controller as the threshold, when the error comes into the scope of the fuzzy controller, the shared controller will automatically switches to the fuzzy controller, switching more smoothly.

The second:

When the driving deviation is not large, the output of the shared controller is as close as possible to the brain control signal to respect the driver's intention. When the deviation is large, the shared controller outputs an optimized steering wheel angle. By setting a cost function, one part ensures that the output of the shared controller is as close as possible to the brain control signal, while the other part ensures that the steering wheel angle of the output will not change too dramatically, making the output angle change more smoothly.

## 5 The simulation analysis of brain controlled vehicle

In the simulation experiment, running-shaped road is used as the virtual experiment scenario, as shown in figure 5. This section is composed of two straight roads and two arc roads, with a linear distance of 200m from the starting point to the first arc, with a length of 157m for each arc and a width of 8.2m for the road. The road is used to test the control characteristics of brain-controlled vehicles in straight lines and corners. FIG. 6 and 7 respectively show the comparison of trajectories of the vehicle under the control of the auxiliary controller and the vehicle under the control of the brain control signal alone. Table 2 shows the comparison between the control effect of the group with the best control effect of brain control alone and that with the addition of auxiliary controller.

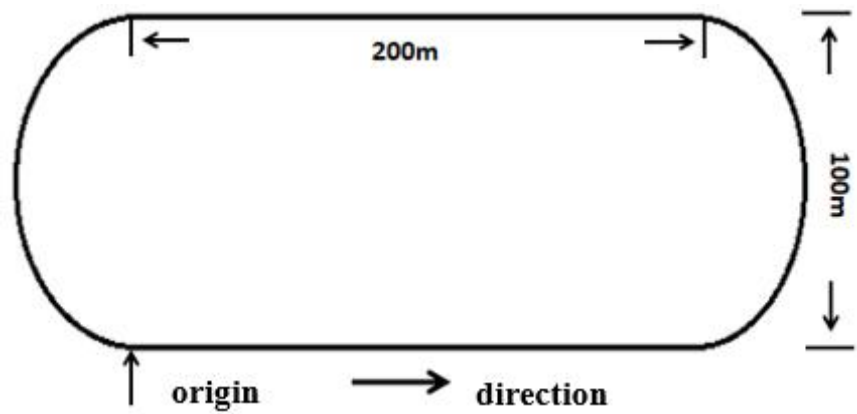

Figure 5 simulation experiment road

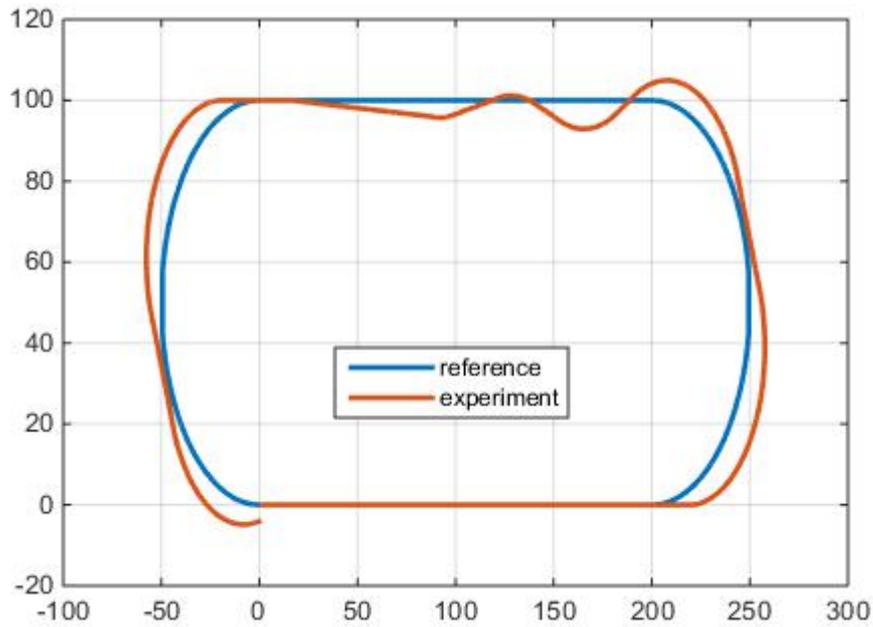

Fig. 6 locus diagram of brain controlled vehicle without auxiliary controller

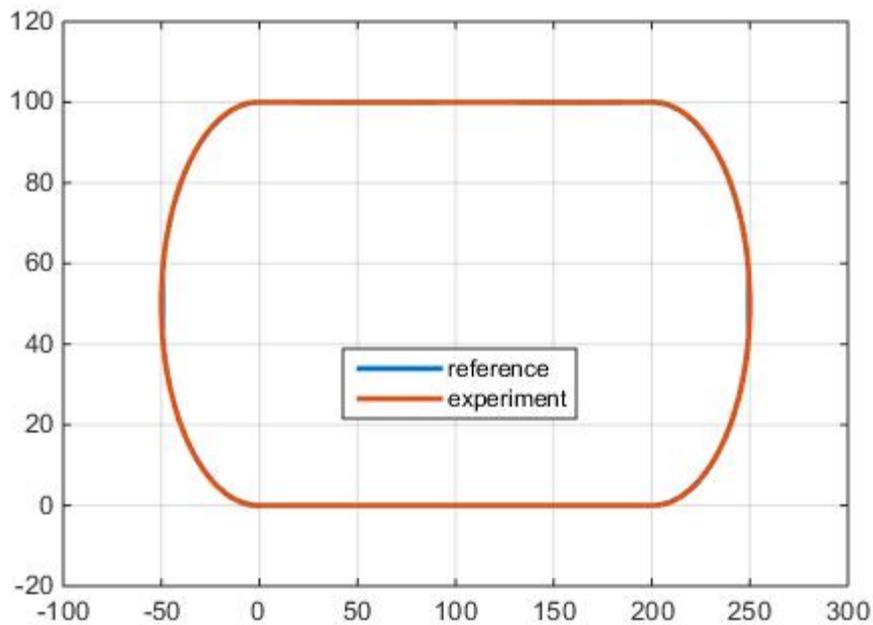

Fig. 7 trajectory chart of brain controlled vehicle with auxiliary controller

Table 2 control effect comparison table

| Control method | Error threshold /m | rotation angle changes of each steering wheel/deg | Brain control regulation times | Trajectory error /m |
|---|---|---|---|---|
| Brain control | 1 | 75 | 9 | 1.3791 |
| Brain control+ auxiliary controller | 1 | 75 | 4 | 0.0588 |

It can be clearly seen that after the addition of the auxiliary controller, the experimental trajectory almost coincides with the set trajectory, and the control effect is significantly improved. Moreover, the adjustment error is greatly reduced, and the control accuracy is significantly improved. The number of brain control adjustment decreased slightly and the driver's burden decreased correspondingly. The results show that the proposed shared control scheme not only reduces the burden of the driver but also improves the control performance of the system.